\documentclass[useAMS,usenatbib]{mn2e}
\usepackage{epsfig}

\def\msun{{\rm {M}}_{\odot}}
\newcommand{\etal}{{et al.}~}
\newcommand{\eg}{{e.g.~}}
\newcommand{\ie}{{i.e.~}}

\def \ltsima{$\; \buildrel < \over \sim \;$}
\def \simlt{\lower.5ex\hbox{\ltsima}}            
\def \gtsima{$\; \buildrel > \over \sim \;$}
\def \gtsima{\mbox{$\; \buildrel > \over \sim \;$}}
\def \simgt{\lower.5ex\hbox{\gtsima}}            

\newcounter{cureqno}

\title[First Stars]{The First Generation of Star-Forming Haloes}
\author[Reed \etal] {
Darren S. Reed$^{1}$ \thanks{Email:
d.s.reed@durham.ac.uk}, 
Richard Bower$^{1}$, 
Carlos S. Frenk$^{1}$,
Liang Gao$^{2}$,
\newauthor
~~~~~~Adrian Jenkins$^{1}$, 
Tom Theuns$^{1,3}$,
and Simon D. M. White$^{2}$\\
$^1$Institute for Computational Cosmology, Dept. of Physics, 
University of Durham,  South Road, Durham DH1 3LE, UK\\ 
$^2$Max Planck Inst. f\"ur Astrophysik, Karl-Schwarzschild Strasse
1, Garching, Munich, D-85740, Germany\\
$^3$Dept. of Physics, Univ. of Antwerp, Campus Drie Eiken, 
Uviversiteitsplein 1, Antwerp, Belgium}

\pagerange{\pageref{firstpage}--\pageref{lastpage}}
\pubyear{2003}

\begin{document}

\maketitle

\label{firstpage}

\begin{abstract}

\noindent We model gas cooling in high-resolution N-body simulations
in order to investigate the formation of the first generation of
stars. We follow a region of a Lambda Cold Dark Matter ($\Lambda$CDM) universe especially
selected to contain a rich cluster by the present day.  The properties
of the dark haloes that form in these sub-solar mass resolution
simulations are presented in a companion paper by Gao et al. The first
gas clouds able to cool by molecular hydrogen line emission collapse
at extremely high redshift, ${\rm z} \approx47$, when the mass of the
dark halo is 2.4${\rm \times 10^5 h^{-1}M_{\odot}}$. By ${\rm z
\approx 30}$, a substantial population of haloes are capable of
undergoing molecular hydrogen cooling although their ability to form
stars is dependent on the efficiency of feedback processes such as
dissociating Lyman-Werner radiation. The mass of the main halo grows
extremely rapidly and, by ${\rm z\approx36}$, its virial temperature
has reached 10$^4$K, at which point gas cooling becomes dominated by
more effective atomic processes.  By ${\rm z \approx 30}$, a small
``group'' of such potential galaxies will have formed unless prevented
from doing so by feedback processes.  By this redshift, massive
($\simgt 100 \msun$) population III stars are able to ionise gas well
beyond their own host halo and neighbouring HII regions can percolate
to form an ionised superbubble. Such patches would be too widely
separated to contribute significantly to reionisation at this time.
The large number density of early cooling haloes in the pre-reionised
universe raises the exciting prospect that this ultra-early generation
of stars may be observable as gamma-ray bursts or supernovae.

\end{abstract}


\section{Introduction}

The first stars in the universe are believed to have formed from
primordial metal free gas in haloes with virial temperatures less than
the $\sim$ 10$^{4}$K threshold for atomic hydrogen line cooling (see
reviews by Bromm \& Larson 2004; Ciardi \& Ferrara 2005 and references
therein).  Primordial gas in these small haloes is instead able to
cool by molecular hydrogen transitions, a less effective process
(Saslaw \& Zipoy 1967; Peebles \& Dicke 1968).  Molecular hydrogen is
produced in a two step reaction chain in which free electrons left
over from recombination serve as a catalyst via H$^{-}$ production
(when z $\simlt$ 100).  Metal free (population III) stars are the
first potential producers of UV photons that can contribute to the
reionisation process, and are the first producers of the metals
required for the formation of population II stars.

Tegmark \etal (1997) developed analytic methods to model early
baryonic collapse via H$_{2}$ cooling.  The early stages of primordial
star formation were later directly modelled in high-resolution
hydrodynamic numerical simulations by Bromm, Coppi \& Larson (1999,
2002) and Abel, Bryan \& Norman (2002).  Yoshida \etal (2003) further
utilised simulations to develop a semi-analytic model based on the
Tegmark \etal methods and included the effects of dynamical heating
caused by the thermalisation of kinetic energy of infall into a
deepening halo potential.  This heating effect varies from halo to
halo, depending on mass accretion rates, and has the potential to
delay baryonic collapse.

Determining the first epoch of baryonic cooling in high redshift
haloes thus requires the use of numerical simulations to follow their
growth.  In this paper, we study the growth of structure in a region
selected to be the site where a rich cluster forms by ${\rm z =0}$ in
a large cosmological simulation. We repeatedly ``resimulate'' this
region at increasing resolution and redshift. In this way, we are able
to model the site of an extremely early generation of star
formation. The simulations employed here are those described in a
companion paper (Gao \etal 2005; G05 hereafter).

We estimate baryonic cooling rates using similar methods to those
described by Yoshida \etal (2003), who performed a detailed
hydrodynamic simulation of a 1 h$^{-1}$Mpc volume in order to
determine under what conditions gas in low mass haloes was able to
cool sufficiently to allow collapse.  Our use of a base region almost
500 h$^{-1}$Mpc on a side allows us to include the important effects
of large-scale modes (White \& Springel 2000; Barkana \& Loeb 2004),
and enables us to model a much earlier generation of baryonic cooling
in rare haloes with low comoving number density.  In addition, we
model H$_{2}$ cooling by explicitly integrating the time-dependent
H$_{2}$ production rate as described in \S~3.

Population III stars are generally expected to be massive ($\sim 100
\msun$) due to the large Jeans mass ($\sim$ 1000$\msun$) during
initial baryonic collapse, and to the inability of zero metallicity
gas clouds to subfragment into small masses during collapse (\eg
Bromm, Coppi \& Larson 1999, 2002; Nakamura \& Umemura 2001; Abel
\etal 2002; Omukai \& Palla 2003; Bromm \& Loeb 2004).  We will assume
that baryonic cooling leads to a {\it single} massive star per halo in
order to make some predictions about their effects on their
surroundings and their potential observability.  By determining the
mean separation and clustering properties of the first star-forming
haloes, we can infer their contribution to reionisation.

If the haloes hosting Pop. III stars are highly clustered, their
associated HII regions could overlap, creating percolating bubbles of
reionisation and, later, metal enrichment.  Alternatively, if they are
weakly clustered, then reionisation from the first stars would be
confined to isolated HII regions.  HII regions from Pop. III stars
have been studied in the redshift range ${\rm z=10-30}$ using 1-D
codes by \eg Kitayama \etal (2004) and Whalen, Abel, \& Norman (2004)
who find that most UV photons produced by a massive ($\sim 100 \msun$)
star are able to escape the virial radius of small haloes such as we
have simulated here.  However, the extent to which high redshift
ionised regions may overlap with those associated with neighbouring
haloes is not yet established.  If Lyman-Werner radiation or metal
enrichment from the first generation of stars is effective at
suppressing the formation of further Pop.  III stars (\eg Omukai \&
Nishi 1999; Glover \& Brand 2001), overlap will not occur and the
original HII region will simply collapse once the stars that power it
reach the end of their lifetime.

As a first step in distinguishing between these scenarios, our
simulations focus on the largest progenitors at high redshifts of a
present day cluster region.  The structure of this paper is as
follows: The numerical techniques and assumptions of these simulations
are discussed in \S~2.  In \S~3, we describe the criteria required for
baryonic collapse.  In \S~4, we show that the first haloes to collapse
do so at extremely high redshift, and show that after a small redshift
interval, many other nearby haloes also satisfy the collapse
criterion.  In \S~5, we consider the implications of these results; we
discuss potential sources of feedback, and we consider whether this
early generation of star formation is open to observational study.
Our conclusions are summarised in \S~6.  Throughout, we assume a flat
$\Lambda$CDM model with the following cosmological parameters, which
are consistent with the combined WMAP/2dFGRS results (Spergel \etal
2003): matter density, $\Omega_m=0.3$; dark energy density,
$\Omega_\Lambda=0.7$; baryon density, $\Omega_{\rm baryon}=0.04$;
fluctuation amplitude, $\sigma_{\rm 8}=0.9$; Hubble constant $h=0.7$
(in units of 100 km s$^{-1}$ Mpc$^{-1}$); and no tilt (i.e. a
primordial spectral index of 1).

\section{Numerical Techniques}

\subsection{The simulations} We use the
parallel gravity solvers Gadget (Springel, Yoshida, \& White 2001) and
Gadget-2 (Springel 2005) to resimulate the region surrounding a rich
cluster identified in a cosmological simulation of a 479 ${\rm
h^{-1}}$Mpc cube (the VLS simulation of Jenkins et al. (2001) and
Yoshida et al (2001)).  This cluster and its environment were part of
the sample resimulated and analysed by Navarro et al. (2004) and Gao
et al. (2004a,b,c). We identified the largest dark matter halo at
${\rm z=5}$ in this resimulation, and then resimulated it again with
resolution improved by an additional factor of 200. We repeated this
procedure three more times stepping progressively back in redshift at
increasing resolution.

Table 1 lists all our simulations, giving their initial and final
redshifts, the mass of an individual particle in the highest
resolution region, the mass of the main halo at the final time (the
original cluster in the case of the VLS simulation), and the
gravitational softening employed. The object that we picked at ${\rm
z=5}$ did {\it not} end up as part of the original cluster selected in
the VLS simulation but rather as a nearby cluster of smaller
mass. This accounts for the difference in masses quoted for the final
``main halo'' in the R1 and VLS simulations. The highest resolution
resimulation ends at redshift 49, has a particle mass of 0.55 ${\rm
h^{-1} \msun}$ and includes a high resolution subvolume of side
90h$^{-1}$kpc. We continue to follow the same halo in a resimulation
ending at redshift 29 with particle mass of 29 ${\rm h^{-1} \msun}$.
Initial conditions were created using the CMBFAST transfer function
(Seljak \& Zaldarriaga 1996), extrapolated to small scales by imposing
${\rm P(k) \propto k^{-3}}$.

\begin{table}
\caption{Simulation parameters.  Each simulation goes from redshift
z$_{\rm start}$ to redshift z$_{\rm fin}$.  M$_{\rm part}$ is the
particle mass in the high-resolution region, M$_{\rm halo}$ the mass
of the main halo and r$_{\rm soft}$ the gravitational force
softening. }
\begin{tabular*}{\columnwidth}{@{}llllll}
\hline\hline
    &  z$_{\rm start}$ & z$_{\rm fin}$  &  M$_{\rm part}$ & M$_{\rm halo}$ & r$_{\rm soft}$ \\
    &                  &                & ${\rm h^{-1}\msun}$ & ${\rm h^{-1}\msun}$ & ${\rm h^{-1}pc}$ \\
\hline
R5 & 599 & 49 & 0.545 & 1.2$\times$10$^{5}$ & 4.8 \\
R4 & 399 & 29 & 29 & 5.1$\times$10$^{7}$ & 17 \\
R3 & 249 & 12 &  1.2$\times$10$^{4}$ & 2.0$\times$10$^{10}$ & 150 \\
R2 & 149 & 5 &  2.2 $\times$10$^{6}$ & 3.4$\times$10$^{12}$ & 800 \\
R1 & 39 & 0 & 5.1$\times$10$^{8}$ & 10$^{14}$ & 5000 \\
VLS & 35 & 0 & 6.8$\times$10$^{10}$ & 8$\times$10$^{14}$ & 30000 \\
\hline
\end{tabular*}
\end{table}

\subsection{Halo properties} Our virialized haloes are identified
using the {\it spherical overdensity} (SO) algorithm (Lacey \& Cole
1994), assuming the spherical tophat model (Eke, Cole, \& Frenk 1996)
in which the $\Lambda$CDM virial overdensity, {\it $\Delta_{\rm
vir}$}, in units of the mean density is 178 at high redshifts when
$\Omega_m\simeq 1$.  For a complex mass distribution such as that
expected to arise from the flat fluctuation spectrum of our high
redshift simulations, SO has an advantage over the simpler
friends-of-friends halo identification algorithm (Davis \etal 1985) in
that it is less likely spuriously to link together neighbouring haloes
or to misclassify highly ellipsoidal but unvirialized structures as
haloes.

The temperature of gas in virial equilibrium within a dark matter halo
is given by
\begin{eqnarray}
{\rm
T_{vir} = 1980 ({\mu \over 1.2}) 
({M_{vir} \over 10^{5} h^{-1} \msun})^{2/3}} 
{\rm 
({\Omega \over \Omega(z)}{\Delta_{vir} \over 18\pi^{2}})^{1/3}
({1+z \over 50})K},
\label{tvir}
\end{eqnarray}
\eg Eke \etal (1996), where $\mu$ is the mean molecular weight (in
units of proton mass) and $M_{vir}$ is the mass within the virial
radius, $r_{vir}$. Simulations by Machacek, Bryan, \& Abel (2001) and
Yoshida \etal (2003) confirm that the local gas temperature is
relatively close to the virial temperature of high redshift haloes,
even within the central density peaks.

Throughout the paper, processes occurring within the haloes are
calculated assuming uniform density and temperature within r$_{\rm
vir}$ because they depend in complex ways on mixing processes which
cannot be treated realistically without simulation.  This includes the
assumption that gas is shock-heated upon initial infall.  We neglect
the possibility that some infalling H$_{2}$ may be destroyed during
shock-heating, but note that H$_{2}$ production is rapid enough that
it would be quickly replenished to a level that allows baryonic
cooling.  Moreover, collisional H$_{2}$ dissociation rates are small
below 10$^{4}$K, so any shock heating strong enough to cause
dissociation would be accompanied by ionisation, providing free
electrons that would increase the eventual H$_{2}$ fraction.  Detailed
hydrodynamic simulations are required to model the heating of
infalling gas accurately.  We assume that the baryonic mass fraction
in the halo is equal to the universal mean value.  Lower baryon
fractions are possible due to the suppression of linear baryon
fluctuations on small scales before the decoupling epoch (\eg
Yamamoto, Sugiyama, \& Suto 1998; Signh \& Ma 2002; Yoshida, Sugiyama,
\& Hernquist 2003).  However, even for haloes somewhat below the
linear theory Jeans mass, gravitational infall into non-linear dark
matter halo potential wells allows the baryons to largely catch up to
the dark matter (\eg Tegmark \etal 1997).  Yoshida \etal (2003) were
able to show that all these approximations are reasonably accurate for
determining whether a halo undergoes sufficient H$_{2}$ cooling for
baryonic collapse of primordial haloes to occur.

One inevitable uncertainty involves the masses and number of the
objects formed in the collapse.  In order to follow the evolution of
the baryonic component to the point at which individual stars are
formed, fully three dimensional simulations with radiative transfer
would be required.  Such simulations are currently beyond the limit of
numerical simulation techniques (Abel \etal 2002; Bromm, Coppi \&
Larson 2002; see Bromm \& Larson 2004 for a review of recent
progress). We thus assume a fiducial collapse model in which the
baryonic cooling of a halo results in a {\it single} $\sim$120$\msun$
star.

\section{Star formation criteria}

We apply the methodology of Tegmark et al.\ (1997) and Yoshida et al.\
(2003) to determine if the baryonic content of a halo is able to
collapse and form stars.  The two requirements are: {\bf (1)} the
cooling rate due to molecular hydrogen must be fast enough for the
halo to cool within a Hubble time; {\bf (2)} the dynamical heating
rate due to mass accretion must be slower than the baryonic cooling
rate.  We assume that efficient baryon cooling leads to the formation
of an increasingly dense gas core that eventually becomes Jeans
unstable (\eg Abel, Bryan, \& Norman 2000).

\begin{figure}
\begin{center}
\epsfig{file=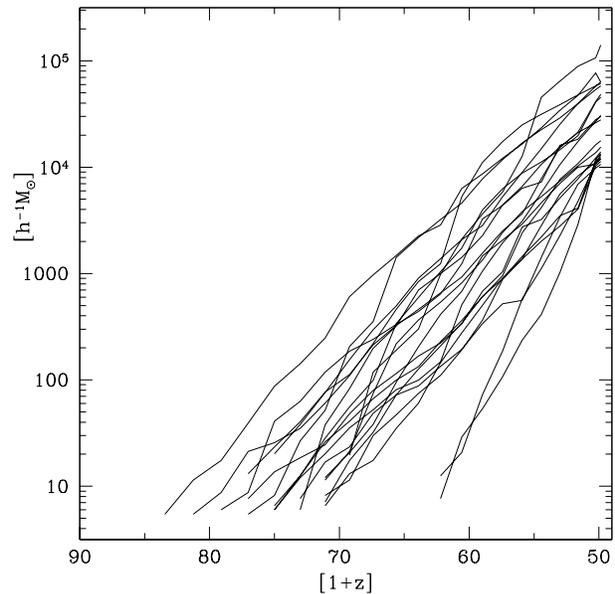, width=\hsize}
\caption{Mass growth of the largest 20 haloes identified in simulation
R5 at ${\rm z=49}$. The mass plotted is that of the largest progenitor
identified at each time.}
\label{trees}
\end{center}
\end{figure}

Fig. \ref{trees} shows that halo growth-rates are remarkably fast, the
haloes typically doubling in mass over a redshift interval ${\rm
\Delta z \simeq 2}$.  Here, the growth of the 20 largest haloes in
simulation R5 is illustrated by tracking the evolution of their
largest progenitor.  We show later that because of the rapid halo
growth, the associated dynamical heating causes a delay in the
baryonic collapse for most haloes. Gas cooling is assumed to depend
purely on the molecular hydrogen abundance, which we calculate by
integrating the density and temperature-dependent H$_{2}$ production
rate over the lifetime of each halo, as we now discuss.

The amount of molecular hydrogen in each dark matter halo is computed
according to the prescription of Tegmark \etal (1997).  The primary
molecular hydrogen production reaction, ${\rm H + H^{-} \rightarrow
H_{2} + e^{-}}$, requires the presence of free electrons to form
H$^{-}$.  The H$_{\rm 2}$ production rate, ${\rm {df_{H_{2}}/dt}
~{\tilde{\propto}}~ T^{0.88}}$ (where ${\rm f_{H_{2}} =
n_{H_{2}}/n_H}$ and ${\rm n_H=n[H] + n[H^+] + 2n[H_2] + n[H^-] \simeq
n[H]}$) is calculated as
\begin{equation}
{\rm {df_{H_{2}} \over dt} \simeq k_mn_{\rm H}x_e },
\label{dfh2dteqn}
\end{equation}
where ${\rm k_m}$ is the production rate, which is limited mainly by
the supply of free electrons for the reaction ${\rm H + e^{-}
\rightarrow H^{-}} + h\nu$.  H$_{2}$ production depends on the density
and temperature of the gas, which are taken to be the volume-averaged
virial quantities, as well as on the free electron fraction x$_{e}$.
We use $x_{e} \equiv x_0 = 10^{-4}$, as obtained for a halo that
virializes at z $\sim$ 50 from the tophat collapse approximation
described in Tegmark et al.

We calculate f$_{\rm H_{2}}$ using a time-integrated production rate.
Other authors have taken a different approach, assuming that the
production of H$_{2}$ stalls at a temperature dependent value when the
remaining free electrons have largely recombined (\eg Tegmark \etal
1997; Yoshida \etal 2003) but, as we show in Appendix A, our haloes
grow too quickly to reach this state.  Since halo growth is so rapid,
we assume that infalling gas mixes uniformly with existing baryons in
the halo, and this has two primary effects.  The first is that the
H$_{2}$ within a halo is diluted by infalling material. The second is
that electrons depleted through recombination are quickly replenished
by infalling gas such that ${\rm x_{e} \simeq x_{0}}$.  The production
of ${\rm H_{2}}$ in a halo is then described by
\begin{equation}
{\rm {dM_{H_{2}} \over dt} \simeq {2\Omega_{H} \over \Omega_{m}}
\left[{dM \over dt}  f_{H_{2, 0}} + {<M{df_{H_{2}} \over dt}>}\right]},
\end{equation}
where ${\rm <M{df_{H_{2}} \over dt}>}$ is the average over adjacent
simulation outputs t$_{i}$ and t$_{i-1}$, ${\rm \Omega_{H}}$ is the
hydrogen density in units of the critical density, and $M$ is the halo
virial mass.  The primordial H$_{2}$ fraction, ${\rm f_{H_{2, 0}}}$ is
quite small $\sim$ 10$^{-6}$ due to H$_{2}^{+}$ photodissociation by
CMB photons at high redshift (Anninos \& Norman 1996; Galli \& Palla
1998; Nishi \& Susa 1999), and so can be ignored.  The H$_{2}$
abundance at time t$_{i}$ is thus:
\begin{equation}
{\rm f_{H_{2, t_{i}}} \simeq { { f_{H_{2, t_{i-1}}}M_{t_{i-1}} + <M{df_{H_{2}} \over dt}>
\Delta{t} }
\over { M_{t_{i}} } }   }.
\end{equation}

\begin{figure}
\epsfig{file=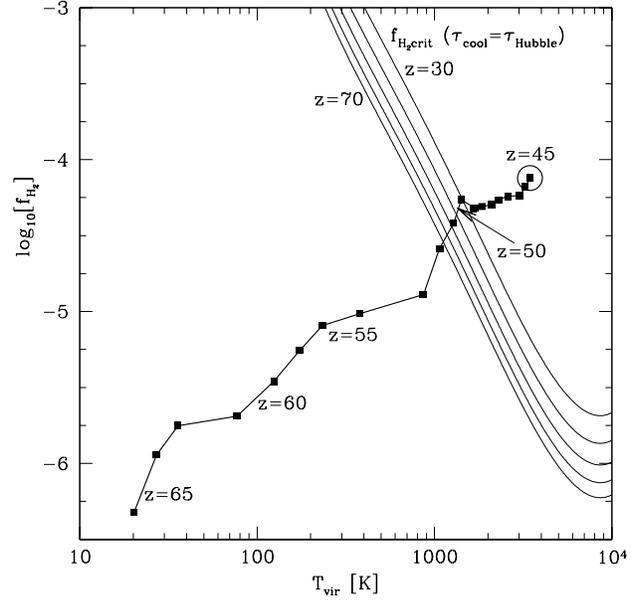, width=\hsize}
\caption{ Evolution of f$_{\rm H_{2}}$ plotted as a function of
T$_{\rm vir}$ (redshift decreases from left to right) for the largest
halo, and compared with f$_{\rm H_{2}, crit}$, which is the H$_{\rm
2}$ fraction required for the baryonic cooling time to be shorter than
the age of the universe.  f$_{\rm H_{2}, crit}$, is shown by {\it
solid curves} for redshift 70, 60, 50, 40, and 30. To be able to cool,
the halo must lie above these curves.  The {\it solid line with data
points} shows f$_{\rm H_{2}}$ computed by integrating ${\rm
df_{H_{2}}/dt}$, and assuming that the H$_{\rm 2}$ abundance is
diluted by infalling material that is both pristine and mixed
uniformly into the halo (see Sect. 3).  In all cases, it is assumed
that infalling gas is shock-heated such that ${\rm T_{gas}=T_{vir}}$.
}
\label{fh2crit}
\end{figure}

We consider a halo to be a star-forming candidate when the molecular
hydrogen fraction becomes large enough that its baryons can cool
within a Hubble time:
\begin{equation}
{\rm \tau_{cool, H_{2}} = \left[{1 \over \gamma-1}{k_{B}T \over
\rho\Lambda(T)}\right] < \tau_{Hubble}}, 
\label{fh2criteqn}
\end{equation}
where $k_B$ is Boltzman's constant and $\gamma$ is the adiabatic index
of the gas. The cooling time, ${\rm \tau_{cool, H_{2}}}$, is
calculated using the cooling function, ${\rm \Lambda(T)}$, from Galli
\& Palla (1998) for molecular hydrogen rotational line transitions.
Fig. \ref{fh2crit} compares the evolution of ${\rm f_{H_2}}$ with the
critical abundance, ${\rm f_{H_2, crit}}$, required for cooling in the
Hubble time. Sufficient H$_{\rm 2}$ for baryonic cooling is produced
at ${\rm z} = 50$ if we assume dilution by newly-added gas, at which
point the halo has a mass of $10^{5} \msun$.  However, we will now
demonstrate that the dynamical heating due to increasing halo mass and
the corresponding increasing virial temperature delays collapse.

\begin{figure}
\begin{center}
\epsfig{file=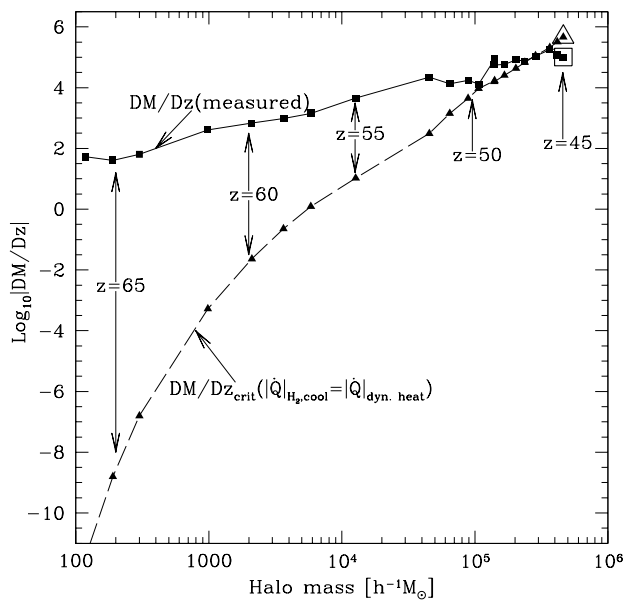, width=\hsize}
\caption{Mass growth rate of the largest halo (DM/Dz, {\it solid
line}) compared to the critical growth rate (DM/Dz$_{\rm crit}$, {\it
dashed line}).  DM/Dz$_{\rm crit}$ is defined as the growth rate for
which the dynamical heating rate exceeds the H$_{\rm 2}$ cooling rate
(Eqn.  \ref{dmdzcriteqn}). The halo growth rate must be below
DM/Dz$_{\rm crit}$ for the baryons to experience nett cooling. The
large square (triangle) denotes measured (critical) mass growth rate
at ${\rm z=45}$.  H$_{2}$ cooling is outpaced by dynamical heating in
this halo until its mass reaches ${\rm 2.4 \times 10^{5} h^{-1}
\msun}$ at ${\rm z=47}$.  }
\label{dmdzcrit}
\end{center}
\end{figure}

Our second criterion for collapse is that the rate of dynamical
heating must be lower than the rate of H$_{2}$ cooling.  If the
dynamical heating rate is larger than the molecular hydrogen cooling
rate, then nett cooling will not occur and baryonic collapse will be
prevented.  Following Yoshida \etal (2003), we require that:
\begin{equation}
{\rm \left| {dQ_{H_{2},cool} \over dt} \right| >
{\rm \left| {dQ_{dyn. heat} \over dt}\right| }},
\label{dmdzcriteqn}
\end{equation}
where Q is the thermal energy per hydrogen atom.  The dynamical
heating rate is given by
\begin{equation} {\rm {dQ_{dyn. heat} \over dt} = {dQ \over dT}{dT \over dt}
= {k_{B} \over \gamma-1}{dT \over dt}},
\end{equation} and the ${\rm H_{2}}$ cooling rate is  
\begin{equation} {\rm {dQ_{H_{2},cool} \over dt} \simeq \Lambda(T)
f_{H_2} n_{H}}. 
\end{equation}
The quantity ${\rm dQ_{dyn. heat}/dt}$ is calculated by tracking the
virial temperature (estimated using Eqn. \ref{tvir}) of the most
massive progenitor halo through the merger tree.  The most massive
progenitor is defined as the largest halo at the previous timestep for
which at least half the mass {\it and} the most bound particle are
part of the target halo.  In Fig. \ref{dmdzcrit}, the dynamical
heating rate falls below the H$_{\rm 2}$ cooling rate after f$_{\rm
H_{2}, crit}$ is reached, at a redshift of 47 when the halo mass is
${\rm 2.4 \times 10^{5} h^{-1} \msun}$.  This suggests that baryonic
collapse in the first halo is delayed by ${\rm {\Delta}z=3}$, or 4
Myr.  Note that there is some uncertainty in the redshift of collapse
due to the large variations of DM/Dz, and also due to the fact that
the halo mass is not well defined by the SO identification when halo
shapes are non-spherical.

\section{Baryonic Cooling in the Simulation}

\subsection{The most massive halo}

Summarising our main conclusions from the previous figures, the most
massive halo in our simulation meets all the criteria for baryon
collapse at $z \simeq 47$. At this point, the halo mass is ${\rm 2.4
\times 10^{5} h^{-1} \msun}$ and has virial temperature of $2300$K.
In this case, dynamical heating of the halo has delayed collapse from
redshift 50 to redshift 47, when the universe is 51 Myr old.  The halo
is then expected to undergo a runaway collapse that results in the
formation of the first star or stars.  Simulations by Bromm \& Loeb
(2004) suggest that a collapsing primordial star may be able to grow
to 120$\msun$ in as little as 10$^{5}$ yr.  Assuming a mass of $\sim$
120 $\msun$, the first supernova could happen $2.5 \times 10^{6}$
years (Schaerer 2002) after the onset of collapse, by redshift 45.
Note that the fiducial stellar mass is just below the predicted range
for pair instability SNe (140-260$\msun$), but is still expected to
undergo an energetic ($\simgt$10$^{51}$ ergs) envelope ejection via
pulsational pair instability, ultimately forming a black hole (Heger
\& Woosley 2002).

\subsection{Smaller haloes}

\begin{figure}
\begin{center}
\epsfig{file=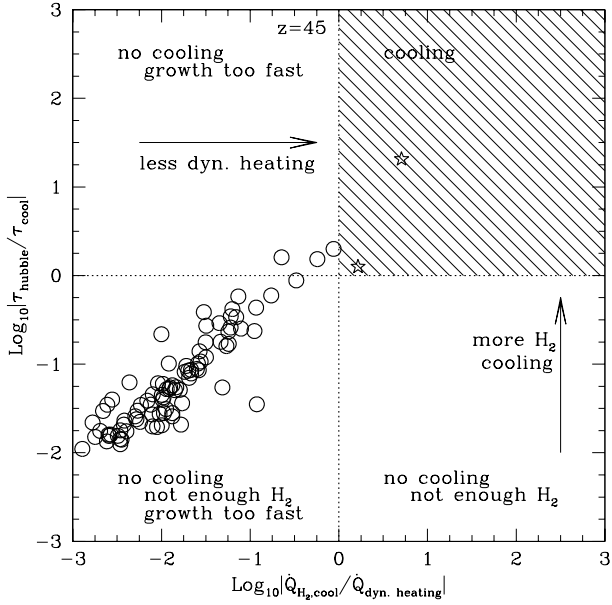, width=\hsize}
\caption{The two criteria for baryonic cooling, ${\rm
\tau_{Hubble}/\tau_{H_{2,cool}} > 1}$ (Eqn. \ref{fh2criteqn}) and
${\rm |\dot{Q}_{H_{2,cool}}/\dot{Q}_{dyn. heat}| > 1}$
(Eqn. \ref{dmdzcriteqn}), for the 100 most massive haloes at redshift
45.  Haloes in the upper right quadrant satisfy both cooling criteria
and are thus expected to undergo baryonic collapse.  Haloes in the
upper half have enough H$_{2}$ such that their baryonic cooling time
is less than the age of the universe. Haloes in the right half have
H$_{2}$ cooling rates that are larger than their dynamical heating
rates and thus experience nett cooling.  Dynamical heating delays
baryonic collapse (haloes in upper left panel) such that only 2 haloes
are capable of cooling by redshift 45, even though 5 haloes have
sufficiently short H$_{2}$ cooling times.  }
\label{quadz45}
\end{center}
\end{figure}

\begin{figure}
\begin{center}
\epsfig{file=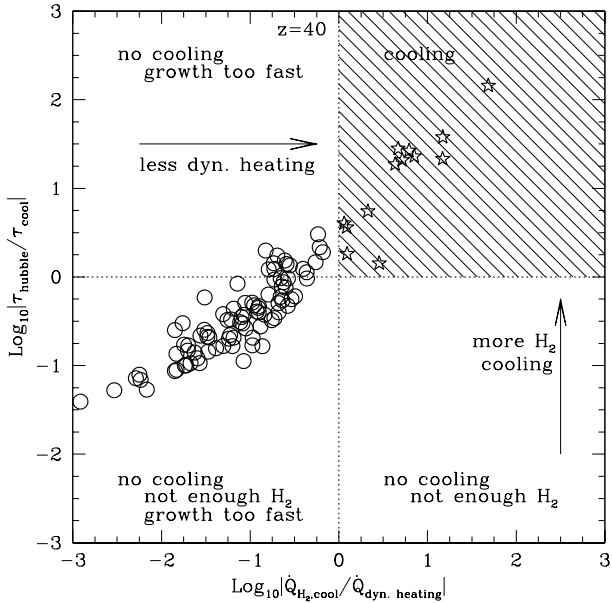, width=\hsize}
\caption{As Fig.~\ref{quadz45}, but for ${\rm z=40}$.  Now, 13 haloes
meet both cooling criteria.}
\label{quadz40}
\end{center}
\end{figure}

\begin{figure}
\begin{center}
\epsfig{file=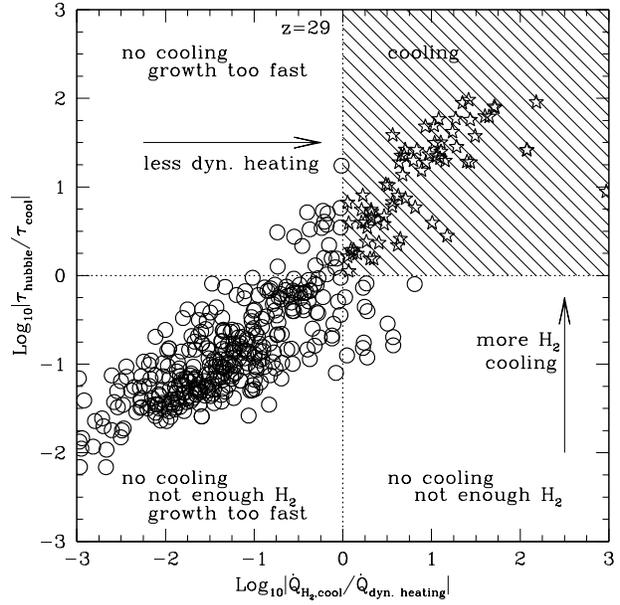, width=\hsize}
\caption{As Figs.~\ref{quadz45} and~\ref{quadz40} but for the 500
largest haloes present at ${\rm z=29}$.  Here, 80 haloes meet both
cooling criteria.}
\label{quadz29}
\end{center}
\end{figure}

In this section, we contrast the cooling of gas in the largest halo
with subsequent cooling in other large haloes in the surrounding
region.  Fig. \ref{quadz45} compares the properties of the 100 most
massive haloes in the simulation at ${\rm z=45}$ with the two criteria
required for gas to cool and collapse. Although 5 of the largest
haloes have sufficient ${\rm H_{2}}$ for rapid cooling at redshift 45,
their dynamical heating rates are still so high that only 2 haloes are
able to cool by this time.  The mass of the smaller of these 2 haloes
is $\approx 10^5$h$^{-1}\msun$.  All of these and 8 other haloes meet
both baryonic cooling criteria by redshift 40, and are thus able
finally to cool, as reflected in Fig.~\ref{quadz40}.  At redshift 40,
the characteristic collapse mass for cooling haloes, which we define
as the mass above which half of the haloes meet both baryonic cooling
criteria, is approximately 2$\times$$10^5$h$^{-1}\msun$.  This
characteristic mass increases slowly to $\sim$
3$\times$$10^5$h$^{-1}\msun$ by ${\rm z=29}$, at which time 80 haloes
in the simulation are capable of cooling (Fig.~\ref{quadz29}).  Most
haloes in Figs. \ref{quadz45}-\ref{quadz29} lie along a rough path
that follows a general sequence of increasing mass from lower-left to
upper-right covering $1.5\times10^4 - 5\times10^5$h$^{-1}\msun$ in
Fig. \ref{quadz45} and $6\times10^4 - 6\times10^7$h$^{-1}\msun$ in
Fig. \ref{quadz29}.  Collapse is delayed by dynamical heating for the
majority of our haloes, as evidenced by the fact that the path of the
mass sequence crosses the ${\rm \tau_{Hubble}/\tau_{H_{2,cool}} = 1}$
line before reaching the ${\rm
\dot{Q}_{H_{2,cool}}/\dot{Q}_{dyn. heat} = 1}$ line.

\subsection{Atomic hydrogen cooling}

Cooling by atomic hydrogen is possible as early as redshift 36 when
the most massive halo reaches a virial temperature of $10^{4}$K, $1.2
\times 10^{7}$ years after it formed its first star.  At this point,
if gas has not all been expelled by prior energy injection, for
example from a supernova, the halo should experience much more rapid
baryonic cooling.  If the high efficiency of metal-free atomic cooling
leads to efficient star formation, this could lead to the formation of
an extremely early galaxy-like system.  The star formation efficiency
and stellar IMF will depend on the details of the molecular cooling
(or dust and metals if the gas is previously enriched) required to
cool the gas below 8000K, although isothermal contraction could also
lead to star formation (Omukai 2001).  By redshift 29, 13 haloes are
hot enough for atomic cooling to occur, and a small ``group'' of
galaxies may already have formed.  Note that this would occur well
before the first stars form in the simulations by Abel \etal (2002),
Bromm, Coppi \& Larson (1999, 2002), and Yoshida \etal (2003), but it
is critically dependent on what happens to the gas after the formation
and evolution of the first stars.

\subsection{Abundance and spatial distribution of the first-star forming haloes}

We cannot use the simulation to determine the abundance of the first
star-forming haloes directly because the high-resolution subvolume
encompassed only 90h$^{-1}$kpc (comoving) in diameter.  The Press \&
Schechter (1974; P-S) prediction for the number density of objects
more massive than $2.4\times 10^{5} h^{-1}\msun$ at ${\rm z=47}$ is
${\rm n_{P-S}(>m) \sim 6 \times 10^{-4} h^3 Mpc^{-3}}$, while the
Sheth \& Tormen (1999; S-T) formulae give a much larger value of ${\rm
5.4 \times 10^{-2} ~h^3 Mpc^{-3}}$.  We do not know {\it a priori}
which of these should be preferred since neither formula has been
tested against N-body simulations in the regime that is relevant here.
Reed \etal (2003) found the S-T mass function to be reasonably
accurate even for rare massive objects at z $=$ 10, and Springel \etal
(2005) showed that the S-T formula is much more accurate than the P-S
formula out to similar redshift.  However, these simulations are still
orders of magnitude away from probing the mass range relevant to this
study. Nevertheless, in what follows we will adopt the S-T
predictions, but the reader has been warned that this extrapolation is
uncertain. Somewhat surprisingly, G05 found that the extended
Press-Schechter theory (Bower 1991; Bond \etal 1991) in the form
worked out by Mo \& White (1996) does accurately reproduce the
abundance of haloes in the rare high density regions tracked in our R4
and R5 simulations over the range ${\rm 30<z<50}$.

It is interesting to consider how rare is the initial overdensity that
led to the halo where baryons were first able to cool in our
simulation.  To estimate this, we identify the Lagrangian volume
${\cal V}$ of the halo by tracing the particles that make up the halo
at ${\rm z=45}$ (which would correspond to the lifespan of a $\sim$
100$\msun$ star formed at z $=$ 47) back to the starting redshift,
${\rm z_{start}}$, of the simulation. The Lagrangian volume ${\cal V}$
of such a rare halo is nearly spherical because it corresponds to a
high peak in the initial density field (e.g. Bardeen \etal
1986). Therefore we can characterise ${\cal V}$ by its mean tophat
linear overdensity, $\delta$, and radius, $L$. We compute the expected
rms density fluctuations on scale $L$, $\sigma(L)$, from our initial
power-spectrum. At ${\rm z_{start}=599}$, when the fluctuation
amplitude is very nearly Gaussian-distributed, $\delta/\sigma(L)=6.5$,
i.e. the halo forms from a 6.5$\sigma$ fluctuation.

The clustering properties of the first metal-free stars will determine
their role in the reionisation and metal enrichment of the surrounding
gas. If these stars are highly clustered, then their HII regions could
overlap and percolate leading to growing patches of reionisation. If,
on the other hand, they are more uniformly distributed, percolation
will be delayed, and the eventual reionisation will be more uniform.
G05 used the Mo \& White (1996) formalism to estimate spatial
correlation functions at ${\rm z=47}$ for haloes at least as massive
as our largest halo. They found a bias factor of 24 which results in a
comoving correlation length of 2.5 h$^{-1}$Mpc, only slightly smaller
than that of star-forming galaxies today. The comoving abundances
estimated above are lower than those of such galaxies, however, and
the luminosities predicted for our first star systems are very much
lower. As a result, the first light sources are relatively weakly
clustered despite their high bias. Neighbouring haloes that lie close
enough to allow the possibility of overlapping HII regions are
extremely rare at redshift ${\rm z=47}$.  As we discuss next, the HII
regions of haloes in the region we have followed may, nevertheless,
percolate.

\section{Discussion}

We have shown that gas is able to cool and collapse in haloes of mass
2.4${\rm \times 10^5 h^{-1}M_{\odot}}$ already at ${\rm z=47}$, and
that this is followed quickly by the collapse of the gas in less
massive haloes in the surrounding region. In just another 22 Myr, at
${\rm z=36}$, the first haloes able to cool by atomic cooling have
formed. In this section, we consider the impact of the collapse of the
first haloes on their surroundings, focusing in particular on the size
of the ionisation patches that are likely to be formed. We consider
the observable consequences of the formation of the first stars in
these haloes at the end of this section.

\subsection{Ionising radiation from the first stars}

We begin by calculating the role of the first stars in ionising the
surrounding intergalactic medium (IGM).  In the next subsection, we
will consider the effect of the first stars on the star formation rate
in surrounding haloes.  In order to estimate limits on the impact of
this first baryonic collapse, we use the flux and lifespan
calculations for Pop.\ III stars by Schaerer (2002).  We first
establish some fiducial values.  The efficiency of production of
ionising photons peaks at a stellar mass of $120 \msun$.  We
parameterise the total ionising photon flux resulting from the
collapse of the gas in a halo as
\begin{equation}
{\rm  n_{ionise} = 1.1\times10^{64} m_{120} f_{120} },
\end{equation}
where m$_{120}$ is the mass, in units of $120 \msun$, of ionising
stars formed per halo, and f$_{120}$ is the total number of ionising
photons emitted per stellar mass relative to the specific ionising
flux from a $120 \msun$ star.  Since f$_{120}$ is a weak function of
stellar mass, n$_{\rm ionise}$ is a weak function of the actual mass
of individual star(s) formed, and is mainly dependent on the total
mass turned into stars.  Later, we will assume that the ${\rm
2.4\times10^5 h^{-1}\msun}$ dark matter halo (which contains
$3.2\times10^4\msun$ baryons), forms a {\it single} 120$\msun$ star
(\ie m$_{120}$f$_{120}=1$).  The maximum possible size of the HII
region around such a star-forming object is ${\rm
76(m_{120}f_{120})^{1/3}}$ kpc (comoving), if we assume no
recombination (\ie the final number of ionised atoms is equal to the
total number of ionising photons produced by the star) and that the
sphere expands into the mean baryon density.  In practise, the high
densities at such high redshift give rise to high recombination rates
that are likely to limit the HII region significantly.

The size of the ionised region is uncertain for several reasons.
Firstly, the recombination rate depends sensitively on the local gas
density.  For simplicity, we assume that gas traces dark matter, which
is likely to be a good approximation outside the virial radius in the
absence of feedback effects, although there is little reason to expect
this to be correct within the virial radius.  Secondly, a further
complication arises from the effect of Compton cooling on the gas.
The Compton cooling timescale is shorter than the recombination
timescale in low density regions at high redshifts.  This could serve
to quench the growth of ionised regions because the recombination rate
increases as gas cools.  Thirdly, the expected short lifetimes of
these massive stars could result in only a fraction of haloes hosting
their first star at any one time.  However, after the first star's
death, the enhanced electron fraction within its HII region may
promote subsequent star formation, albeit of stars possibly of lower
mass (O'Shea \etal 2005).  Fourthly, supernovae from the first stars
may create bubbles larger than their HII regions.  Finally, as we
discuss in \S~5.2, Lyman-Werner radiation may destroy molecular
hydrogen in neighbouring haloes, thus limiting the number of stars
that form in H$_{2}$-cooled hosts, although stars in atomic-cooled
haloes may soon dominate the stellar population.  In practice, of
course, the ionised region will not be spherical but will depend on
the morphology of the surrounding gas distribution.

We have estimated the size of an HII region using a steady-state 1-D
code. The recombination rate is governed by the rate coefficient,
${\rm k_{recomb}=1.88\times10^{10}T^{-0.64}}$ (cm$^3$s$^{-1}$)
(Hutchins 1976), and the temperature of the ionised gas is set to
${\rm T=10^4}$K.  The size of the HII region is taken simply to be the
radius at which the recombination rate within the sphere is equal to
the ionising photon production rate.  We assume that the gas follows
an NFW profile (Navarro, Frenk \& White 1996, 1997) with a
concentration of 1.75 at ${\rm z=47}$, and 1.87 at ${\rm z=29}$, which
provides a good fit to the dark matter profile in the simulation
(G05).  To estimate the effects of the shock front, we smooth the gas
in the central portion of the HII region to have a uniform density
over the sound crossing distance for a star with a 2.5 Myr lifespan
(assuming ionised gas at 10$^{4}$K).  This smoothing also makes the
estimated HII region size less sensitive to the assumed profile shape.
The mass density surrounding the largest halo in the simulation
approximately follows an NFW profile extrapolation until it reaches
$\sim$ 10 times the mean density at $\sim$ 3${\rm r_{vir}}$, beyond
which we assume a uniform density.

\begin{figure*}
\begin{center}
\epsfig{file=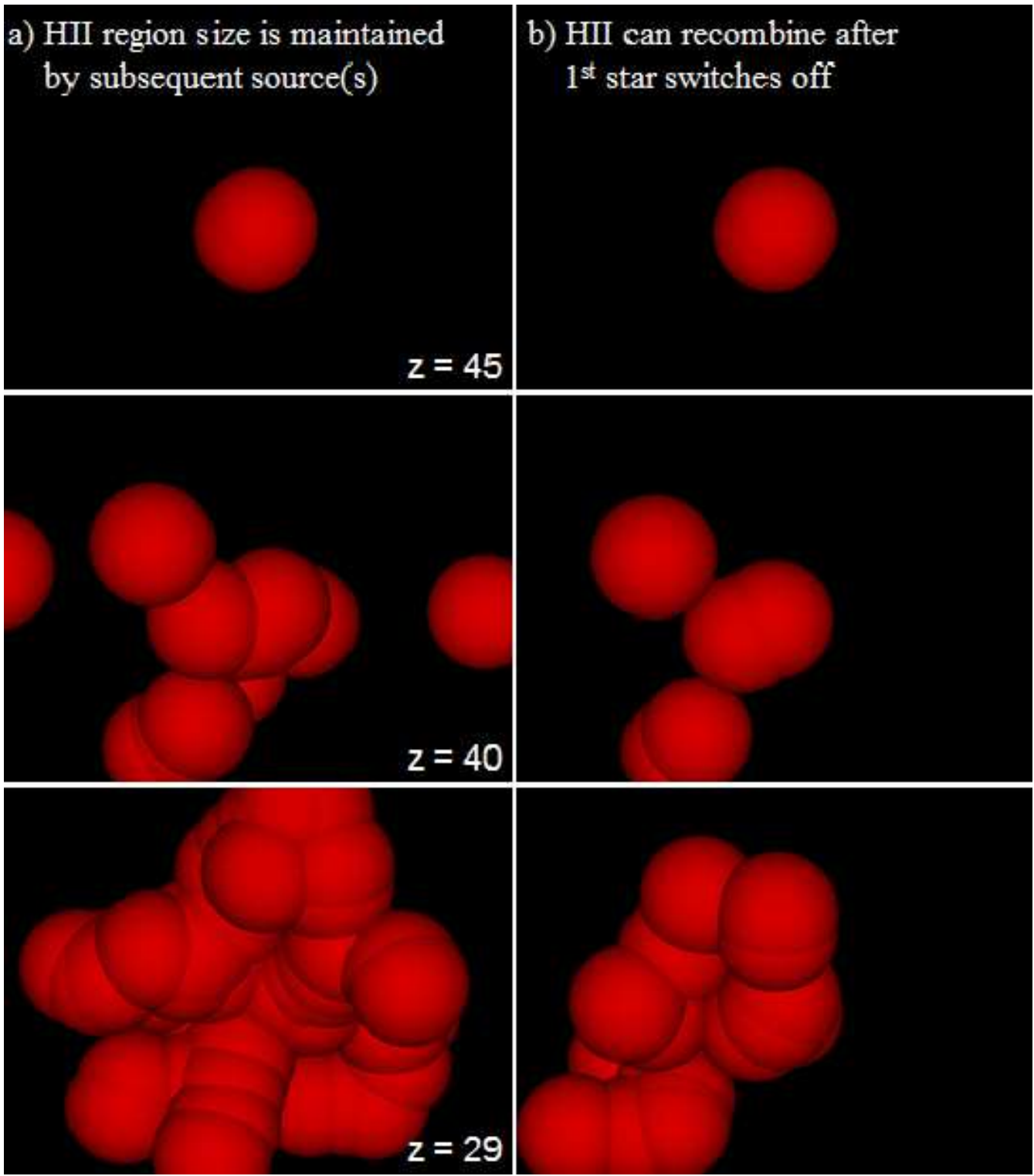, width=\hsize}
\caption{In this schematic image, an HII region powered by a 120
$\msun$ Pop.~III star is centred on each ${\rm z=29}$ R4 halo that
meets the H$_{2}$ cooling criteria of \S~3 at redshifts 45, 40, and
29.  In the left panel, the 21 kpc (comoving) radius HII regions (see
text) are assumed to be maintained after the end of the first star in
each halo by subsequent ionising sources.  In the right panel, HII
regions are assumed to survive for only one recombination time ($\sim$
1 Myr) beyond the life of the first star in each host.  Significant
local percolation of HII regions is seen in both cases.  In \S~5.2, we
discuss feedback mechanisms that may greatly alter the number of
star-forming haloes in regions such as this.  }
\label{hii}
\end{center}
\end{figure*}

For the fiducial m$_{120}$f$_{120}=1$ case, at redshift 47, the HII
region extends just beyond the virial radius, to ${\rm 1.2 r_{vir}}$
in our calculation.  However, by redshift 29, physical densities are
low enough that over half the ionising photons from a single
120$\msun$ star will escape a halo of the characteristic mass for
H$_{2}$-cooled haloes at this redshift (${\rm \sim 3\times10^5
h^{-1}\msun}$).  Additionally, a much larger number of haloes are now
susceptible to H$_{2}$ cooling, raising the possibility that
percolation of HII regions could create a large HII bubble.  At ${\rm
z = 29}$, the HII region around a single 120$\msun$ star would extend
to 9r$_{\rm vir}$ (21 h$^{-1}$kpc, comoving). This is large enough for
the entire $\simeq$ 75 h$^{-1}$kpc (comoving) radius region that we
have simulated to have an HII filling factor of $\sim$ 1, provided
that each halo maintains a similar sized HII region.  Of course, such
regions are rare and the contribution to the ionised volume of the
universe as a whole is small (see \S~5.3).  The percolation of these
HII bubbles is illustrated in Fig. \ref{hii}, left panel.  This
scenario is probably unrealistic due to short recombination times of
$\sim$ 1 Myr in the high density ($\simgt$ 10 times the mean density)
vicinity of these haloes, and due to potential removal of gas by
feedback.  The short recombination times imply that percolation of
ionised bubbles will be significantly reduced if a new ionising
source(s) does not form before the end of the first star's lifetime.
The right-hand panel in Fig. \ref{hii} illustrates how percolation
will be reduced if each HII region survives only one recombination
time after the lifetime of the primary star.  Finally, we have
considered whether Compton cooling could affect the size of individual
HII regions by enhancing the recombination rate, and find that this
effect is likely to be unimportant over the lifetime of a massive
star.  At redshift 29, the 1.4 Myr Compton cooling timescale is of
similar order, but generally longer than the typical local
recombination time.

Thus, while the size of an HII region is quite uncertain, it seems
possible that a large bubble around the first star-forming halo could
become ionised well before the rest of the universe.  One might then
expect similarly large and widely separated HII bubbles to be
scattered throughout the Universe prior to reionisation.

\subsection{Feedback from the first generation of stars}

In \S~4, we neglected the impact of the collapse of the first stars on
neighbouring haloes.  The picture we have presented so far only takes
into account the ability of a halo to cool and form stars according to
local criteria without taking into account external influences.  We
now briefly discuss how negative feedback from the first stars may
impact star formation in the surrounding region.

There are three (possibly more) feedback mechanisms that could mean
that the ability of a halo to cool is not a local process, and does
not depend solely on the mass of the dark matter halo and its
formation history.

\begin{itemize}
\item{\it Lyman-Werner radiation:} Radiation in the Lyman-Werner bands
can dissociate hydrogen molecules.  If the destruction rate exceeds
the rate of production, the halo will not build up the amount of H$_2$
needed to cool.  As a result, the formation of the first star may
``sterilise'' the surrounding region and prevent other stars from
forming (\eg Haiman, Rees, \& Loeb 1997; Ciardi, Ferrara, \& Abel
2000; Haiman, Abel, \& Rees 2000; Ricotti, Gnedin, \& Shull 2002;
Mackey, Bromm, \& Hernquist 2003; Yoshida \etal 2003).

The photo-dissociation timescale for ${\rm H_2}$ is given by
\begin{equation}
{\rm
t_{diss} = k_{diss}^{-1}F_{trans}^{-1}}
\end{equation}
\begin{equation}
{\rm
k_{diss} = 1.1 \times 10^{8} j_{LW} (s^{-1})}
\end{equation}
(Abel \etal 1997), where j$_{\rm LW}$ is the Lyman-Werner flux at
$h\nu=12.87$eV in units of ${\rm 10^{-21} ergs~s^{-1} cm^{-2}\
Hz^{-1}}$.  In the absence of self-shielding, the transmission factor
F$_{\rm trans}$ is equal to 1.

It is convenient to parameterise the dissociation timescale in terms
of the photo-dissociating luminosity (12.24ev-13.51ev) for the
fiducial 120$\msun$ star ${\rm L^{LW}_{120}}$:
\begin{equation}
{\rm
t_{diss} = 1.3~{1 \over L^{LW}_{120}} \left({d \over {1Mpc}}\right)^{2}\left({{50}
\over {1+z}}\right)^{2}~\frac{1}{F_{trans}}~ (Myr)},
\end{equation}
where {\it d} is the distance from the star in comoving Mpc.

In the absence of self-shielding (see below), a single $120\msun$ star
produces enough Lyman-Werner flux to dissociate H$_{2}$ within a
volume of $\sim$ 1 comoving Mpc$^3$.  (Here, the ``sterilisation by
dissociation'' radius is determined by estimating the distance at
which the dissociation timescale is comparable to the H$_{2}$
formation timescale, which is typically $\simgt$ 1 Myr.)

This calculation ignores the fact that the surrounding haloes are
likely to contain dense cores. If the optical depth in these regions
is sufficiently high, the inner mass will escape dissociation because
of self-shielding in the outer parts (\eg Glover \& Brand 2001;
Kitayama \etal 2001; Machacek \etal 2001).  Unfortunately, the
effectiveness of self-shielding depends on the internal velocity
structure of the halo gas caused by infall and turbulent motions. The
properties of these flows are uncertain and they will be difficult to
model adequately even in the highest resolution simulations.

If the gas velocities are negligible, the transmission factor is given
approximately by:
\begin{equation}
{\rm F_{trans} =
min\left[1, \left({N_{H_2} \over 10^{14} cm^{-2}}\right)^{-3/4}\right] }
\label{ss}
\end{equation}
(Draine \& Bertoldi 1996), where N${\rm _{H_2}}$ is the H$_2$ column
density, integrated from outside-in.  Because it neglects gas motions,
this calculation gives the maximum possible self-shielding (Machecek
\etal 2001).  For an NFW halo profile with a concentration of 1.75,
and assuming that the H$_{2}$ density traces the mass density,
Eqn.~\ref{ss} implies that 90$\%$ of the H$_{2}$ mass (${\rm r >
0.2}{\rm r_{vir}}$) will be prone to dissociation by a single
120$\msun$ star located at a distance of 65 kpc at redshift 45, for a
typical ${\rm H_2}$ fraction of 10$^{-4}$.  This implies that
dissociation is still important at large distances.  For the typical
halo separation in our simulations ($\sim$ 10 h$^{-1}$kpc), a
transmission factor of less than 1.5$\times 10^{-4}$ would be required
to ensure nett H$_2$ production (${\rm t_{diss} \simgt 1}$Myr) at
redshift 45.  However, at this distance, the central baryonic mass
that is adequately shielded is much less than that required to form a
massive star unless either the baryon density or the H$_{2}$ fraction
are significantly enhanced.  On the other hand, baryonic cooling is
likely to concentrate baryons strongly towards the halo centre (\eg
Yoshida \etal 2003), increasing the Lyman-Werner optical depth and
decreasing the effect of Lyman-Werner feedback.  These issues can only
be addressed with detailed numerical hydrodynamic simulations.

We estimate that neither L-W shielding by intergalactic H$_{2}$ nor by
intergalactic neutral Hydrogen will be important.  Using the
transmission factor of Eqn.~\ref{ss}, and assuming a primordial
H$_{2}$ fraction of 10$^{-6}$ at redshift 50, L-W shielding by
intergalactic H$_{2}$ will be negligible except over very large
distances (${\rm F_{trans}=0.1}$ at 1~Mpc, comoving).  This estimate
likely over-predicts the inter-galactic shielding effect because
doppler shifts due to Hubble expansion and intrinsic gas motions will
decrease the optical depth.  More importantly, any shielding by
inter-galactic H$_{2}$ will be short-lived because it will quickly be
dissociated.  Shielding by inter-galactic neutral hydrogen is also
ineffective because L-W absorption will be largely confined to the
closely spaced lyman series lines in the upper energy L-W bands.  Even
at a distance of 1 Mpc (comoving), less than half of total L-W
radiation will be absorbed by neutral H at redshift 50.

We have rerun our H${_2}$ production model including the dissociating
effects of Lyman-Werner radiation under the assumption that
self-shielding is unimportant.  The L-W flux from a single 120 $\msun$
star is enough to quickly sterilise the entire (75 h$^{-1}$kpc
comoving radius) high resolution subvolume of the R4 simulation.
However, if we then switch off the L-W flux after a 2.5 Myr stellar
lifetime, a second halo is able to produce enough H$_{2}$ to meet the
baryonic cooling criteria within 1 - 2 Myr.  Thus, the sterilising
effects of a single massive star are temporary, but may nevertheless
limit the star formation density to $\sim$ 1 star per comoving Mpc per
Myr in H$_{2}$-cooled hosts.  However, if self-shielding within dense
baryonic cores can cause a significant time delay between the onset of
baryonic collapse and subsequent sterilisation of neighbouring haloes,
then there may exist a brief window during which multiple haloes will
have time to produce enough H$_{2}$ for baryonic cooling.  A
sufficiently long window could allow a rapid ``mini-burst'' of H$_{2}$
induced population III star formation in a previously sterilised
region.

Thus, depending on the detailed structure of collapsing clouds, the
radius out to which Lyman-Werner feedback is effective could be as
large as $\sim$ 1 Mpc, which would substantially limit the number of
haloes that undergo collapse via H$_{2}$ cooling.  Self-shielding is
likely to mediate this feedback to some degree, but it is probably
unimportant except for dense baryonic cores.  However, we stress that
even if all the H$_2$ in a halo were destroyed by external
Lyman-Werner radiation, atomic cooling alone should still be
sufficient to trigger baryonic collapse in our haloes once they reach
a sufficiently high virial temperature (\eg Omukai 2001).

\item{\it X-ray and far UV radiation:} The first supernovae and
mini-qso's will bathe the surrounding region with ionising photons
(\eg Madau \etal 2004). Those of sufficiently high energy have small
ionisation cross-sections and will escape from the HII
region. However, even if they only ionise a small fraction of the
neutral hydrogen in a nearby halo, the increase in the free electron
density will catalyse the production of H$_{2}$.  If this process is
significant in the centres of these haloes, it would promote the
collapse of their gas, acting as a positive feedback mechanism (\eg
Haiman, Rees, \& Loeb 1996; Haiman, Abel, \& Rees 2000; see however,
Machacek, Bryan, \& Abel 2003, who find that a uniform X-ray
background has little effect on H$_{2}$-induced cooling and collapse
of haloes in simulations).

\item{\it Metal enrichment and SNe:} The energy released by the
explosion of the first SNe sends a blast wave into the surrounding
IGM. If cooling haloes are fragile, the associated mechanical energy
can remove portions of their outer envelope and prevent further
cooling (\eg Rees 1985; Bromm, Yoshida, \& Hernquist 2003; Kitayama \&
Yoshida 2005).

Pop. III stars with masses in a narrow range around 200$\msun$ are
thought to end their life as a pair production supernova (\eg Heger \&
Woosley 2002).  In this case, the supernova ejecta is very metal rich
and, if this material mixes in with the gas in nearby haloes, the
overall gas cooling efficiency will be enhanced.  Metal enrichment is,
in any case, required in order to enhance baryon cooling at low
temperatures ($<$10$^4$K) so as to promote the fragmentation that
results in the low-mass IMF typical of population II stars (Schneider
\etal 2002, 2003; Mackey, Bromm, \& Hernquist 2003).  At redshift
$\sim$ 30, energetic ($\sim 10^{53}$) SNe blastwaves may expand up to
$\sim$ 20 kpc (comoving) before the initial fast expansion is
``stalled'' (estimated according to Yoshida, Bromm, \& Hernquist
2004), suggesting that metal enrichment from neighbouring haloes is
possible.

Thus, supernovae blast waves can produce both positive and negative
feedback.  However, the region affected by these processes is likely
to be small compared to the region potentially affected by the
Lyman-Werner radiation.

\item{\it H$^{-}$ Photo-detachment} Another possible mechanism
suppressing H$_{2}$ formation is the photo-detachment of H$^{-}$ ions
before they can combine with a hydrogen atom to produce H$_{2}$.  We
estimate the radius over which photo-detachment suppresses H$_{2}$
formation by equating the photo-detachment rate to the rate at which
H$^{-}$ reacts to form H$_{2}$ (${\rm H + H^{-} \rightarrow H_{2} +
e^{-}}$; rate taken from Hirasawa 1969), assuming virial overdensity.
We assume the effective temperature and luminosity of a 120 $\msun$
Pop. III star from Schaerer (2002), and employ the Tegmark \etal
(1997) fitting function to photo-detachment cross-sections from
Wishart (1979).  This yields a photo-detachment distance of 2.5 kpc
(comoving) at ${\rm z = 29}$, roughly equal to the virial radius of a
typical halo on the verge of baryonic cooling at this time, and much
smaller than the HII region expected from such a massive star.  Thus
we do not expect H$^{-}$ photo-detachment by ${\rm \sim 100 \msun}$
Pop. III stars to impact baryon cooling in neighbouring halos.

\end{itemize}

{\it In summary}, our calculations show that Lyman-Werner radiation
could ``sterilise'' a region as large as $\sim$ 1 Mpc (comoving)
around a primordial massive (120 $\msun$) star. However, the effects
of self-shielding are so uncertain that we cannot rule out the
possibility that dissociation feedback may be ineffective at
suppressing further star formation.  The uncertainties in other
feedback mechanisms are also considerable. In the preceding section we
discussed the possibility that the HII regions around the first stars
might percolate giving rise to large multi-halo HII bubbles. Whether
or not this happens depends primarily on how effective the
Lyman-Werner sterilisation is at suppressing star formation in the
vicinity of a star, and on the impact of the other feedback schemes
discussed above.  However, even if sterilisation does occur, this can
only delay star formation in the region, since atomic cooling, already
possible in 13 haloes in our simulation by ${\rm z=29}$, will
ultimately become the dominant path for star formation.  For the case
of maximally effective sterilisation, the S-T number density of haloes
capable of atomic cooling surpasses that of H$_2$-cooled haloes
($\sim$ 1 per h$^{-3}$Mpc$^{3}$) by z $\sim$ 25. Gas in atomic-cooled
haloes could dominate the total stellar mass and UV radiation
considerably earlier than this if, as seems likely, this gas turns
into stars more efficiently than gas within H$_{2}$-cooled haloes due
to faster rate and greater temperature sensitivity of atomic cooling.

Unfortunately, it is not yet possible to rule out any particular
scenario.  We proceed by exploring the extreme case where dissociative
feedback is unimportant, and percolation leads to the formation of a
large ionised region.

\subsection{Contribution to reionisation}

The sequence of events following the formation of the first generation
of Pop.~III stars (z $\simgt$ 45) is uncertain because of our poor
understanding of the interplay between negative feedback from
Lyman-Werner radiation and positive feedback from the X-ray/UV
background. If these processes are unimportant or if positive feedback
dominates, our calculations suggest that large ($\sim$ 100 h$^{-1}$kpc
comoving) patches of the universe will be reionised at ${\rm z}>30$.
The first regions of the universe to collapse may thus become
reionised much earlier than the ${\rm z \simeq 20}$ inferred from the
WMAP microwave background radiation data (Kogut \etal 2003). This
could have dramatic effects on low mass galaxies forming in these
regions whose gas could be removed suppressing subsequent star
formation.

Observational constraints derived from observations of high redshift
QSOs (Fan \etal 2001) and from the temperature of the high redshift
IGM (Theuns \etal 2002) imply that complete re-ionisation occurs at
${\rm z=6-9}$. In contrast, the high optical depth for Thomson
scattering inferred from the WMAP data suggests a much earlier epoch
for reionisation, at least over significant fractions of the Universe.
It is interesting to consider whether large patches of early
reionisation could reconcile these contradictory constraints (see \eg
Bruscoli, Ferrara, \& Scannapieco 2002; Cen 2003; Ciardi, Ferrara, \&
White 2003; Haiman \& Holder 2003; Sokasian \etal 2004).  However,
according to the number densities inferred from the S-T formula,
regions such as the one we have simulated are quite rare and so they
can only make a small contribution to global reionisation and to the
Thompson optical depth.  Moreover, stars in H$_{2}$-cooled haloes
could even delay reionisation if their sterilising effects prevent
H$_2$-cooled star formation over large distances.  Alternatively,
strong Lyman-Werner feedback from the first population III stars might
actually expedite reionisation by preventing the expulsion of gas by
feedback associated with star formation in early low mass haloes,
leaving plentiful gas supplies when haloes grow large enough for
efficient atomic cooling.

\subsection{Observability}

Finally, we consider whether the first generation (${\rm z} \simgt
45$) of population III stars that is predicted by our simulations
could actually be observed.  We focus first on the case where H$_{2}$
is unaffected by feedback processes, which are highly uncertain.  The
supernovae resulting from the collapse of these stars are potentially
observable, particularly if they can be detected as gamma-ray bursts
(GBRs).  The time dilation of bursts occurring at higher redshifts
implies that the source can be observed closer to maximum light, thus
compensating for the increasing luminosity distance. In a recent
paper, Gou et al. (2004) showed that GRBs can be detected in X-rays
out to ${\rm z=30}$ and beyond in exposure times as short as 300 sec.

The rest-frame ultraviolet emission from the burst will be strongly
absorbed shortwards of Ly-${\alpha}$, so that detection will need to
be made in the mid infrared. Gou et al show that the James Webb Space
Telescope will have the required sensitivity to detect GRBs out to
${\rm z=37}$ in the $4.8\mu m$ band in only 1 hour.  Detection of GRBs
at higher redshift should also be possible in longer wavelength bands
and larger exposure times.  Radio afterglows may also possible at z
$\sim$ 30 (Ioka \& Meszaros 2005).

The number of haloes massive enough to support star formation lying in
our past lightcone at ${\rm z}>45$ is of the order of 10$^{9}$,
according to the S-T prediction. If we assume that each of these
haloes forms a single massive star with a 2.5 Myr lifespan whose
endpoint is a GRB, then the event rate over the entire sky would be of
the order of one per month.  Our estimated abundances (which, as we
discussed in \S~4.4, are uncertain) are higher than, but in rough
agreement with those of Miralda-Escude (2003) who, on the basis of the
P-S formula, predicts the highest redshift star in our past lightcone
to be at redshift 48, assuming that such stars form in haloes with a
virial temperature of 2000K.  Our z $\sim$ 50 supernovae rates are of
similar order to those predicted by Wise \& Abel (2004) on the basis
that supernovae are hosted by haloes of mass $\sim$ 10$^{5} \msun$
whose abundances is given by the S-T model.  Note that we have ignored
GRB beaming in this calculation; the number of observable GRBs that
are beamed toward us is likely to be lower by a factor $\sim$
100$-$1000 (\eg Frail \etal 2001).

The rate of potentially observable GRBs or SNe forming in haloes where
gas has cooled via H$_{2}$ increases rapidly at lower redshifts.  At
${\rm z = 29}$ in our simulation we expect $\sim$ 10 supernovae per
day over the entire sky, or 1 per $\sim$ 10 square degrees per year
(again, ignoring beaming effects).  If sterilisation by Lyman-Werner
radiation strongly suppresses star formation, the number of observable
GRBs at z$>$29 could be reduced by a significant factor, depending on
the uncertain effects of self-shielding.  However,even in that case,
the number density of haloes capable of sustaining atomic cooling
would soon surpass that of H$_2$-cooled haloes, leaving open the
possibility of abundant pre-reionisation observational targets.  The
increase in the expected event rate at lower redshifts suggests that
GRBs or SNe should indeed be observable in the pre-reionised universe.

We conclude that if the highest redshift Pop.~III stars end their life
as GRBs, the prospects for their detection are promising.  Even if
these stars do not have a GRB phase or if the beaming angle is small,
observational evidence of the first stars may still be available, for
example, through direct detection of SNe (Panagia 2003) or of emission
from an early generation of accreting black holes (\eg Madau \etal
2004).  The difficulty, however, will lie in identifying such rare and
faint objects and distinguishing them from foreground contaminants.

\section{Conclusions}

We have modelled the cooling of gas by molecular hydrogen line
emission in a series of N-body simulations in order to identify the
first generation of dark matter haloes capable of forming stars. We
followed the formation of structure in a region selected to host a
rich cluster today, repeatedly resimulating, at increasing resolution,
the largest progenitors of a massive halo at the final time. To
determine whether the gas is able to cool, we estimated the production
rate of H$_{2}$ and considered the balance between radiative cooling
and dynamical heating.  The dramatic growth of the halo at early times
has the nett effect of delaying, but not preventing the cooling of
gas.  Our main results are as follows.

\begin{itemize}

\item{\bf The first star-forming haloes:} Since we have followed
evolution in a special region of the universe (selected to have a rich
cluster today), we find that formation of the first stars occurs
significantly earlier than commonly found in simulations of volumes
too small to sample such rare overdensities.  At a redshift of 47,
there is already a population of haloes in the simulation that are
undergoing baryonic collapse via H$_2$ cooling and can host the first
stars.

The mass of these haloes is typically 2.5${\rm \times 10^5
h^{-1}M_{\odot}}$ and their comoving space density is comparable to
that of dwarf galaxies today.  By redshift 29, 80 haloes in the
high-resolution simulation subvolume of radius 75h$^{-1}$kpc are
capable of forming stars.  The largest halo reaches a virial
temperature of 10$^4$K, at which gas in equilibrium can begin to cool
by more efficient atomic processes, at ${\rm z=36}$, potentially
giving rise to a first generation galaxy.  By ${\rm z= 29}$, 13 haloes
in the simulation have grown large enough to sustain atomic cooling.

\item{\bf Ionising effects:} The formation of the first stars will
ionise the gas around them. At ${\rm z=47}$, the hydrogen
recombination rates are high enough that these HII regions are likely
to be contained almost entirely within their own host halo.  However,
by ${\rm z=29}$, gas densities have dropped sufficiently that massive
($\sim$ 100 $\msun$), population III stars are able to ionise regions
significantly beyond the halo.  Percolation of neighbouring regions
could lead to the formation of an HII superbubble.  However, such
regions would be too widely separated to contribute significantly to
reionisation at this time.

\item{\bf Feedback:} Feedback effects that could alter the cooling
properties of the gas are likely to occur but their nature (and, in
some cases, even their sign) is difficult to calculate. Lyman-Werner
radiation from just a single massive star could sterilise a large
surrounding volume by dissociating H$_2$, severely limiting star
formation throughout the volume. However, the effectiveness of this
feedback channel would be limited if the H$_2$ within dense halo cores
is self-shielded by the outer halo layers.  Supernova feedback could
hinder star formation by removing gas, or could enhance star formation
by blast wave compression, by metal enrichment, or through high energy
radiation which might produce electrons that catalyse H$_2$ formation.

Even if feedback is effective enough to suppress H$_2$ cooling within
a region of size $\sim$1 Mpc (comoving) around the first star, further
star formation is still expected in haloes where gas can cool by
atomic processes. This happens as early as ${\rm z=36}$ in our
simulated region.

\item{\bf Observability:} The prospects for observing the very
earliest generation of stars are encouraging.  The James Webb Space
Telescope may be able to view directly pre-reionisation supernovae and
gamma-ray bursts from population III objects in the rest-frame
mid-infrared (Panagia 2003).  X-ray emission or radio afterglow from
these early gamma-ray bursts may also detectable (Gou \etal 2004).
The event rate of supernovae at ${\rm z=29}$ and higher could be as
large as $\sim$ 10~per day.  However, if feedback due to Lyman-Werner
radiation is important, the event rate could be substantially lower.

\end{itemize}

\section*{Acknowledgments}
DR is supported by PPARC.  RGB is a PPARC Senior Fellow.  TT thanks
PPARC for the award of an Advanced Fellowship.  The simulations were
performed as part of the Virgo consortium programme, on the IBM
Regatta at Max Planck Inst. f\"ur Astrophysik in Garching, Germany,
and on the Cosmology Machine supercomputer at the Institute for
Computational Cosmology in Durham, England.  We are grateful for
insightful and constructive criticism by the referee, Noaki Yoshida.

{}

\appendix
\section{}

In this appendix, we discuss the production of H$_{2}$, comparing our
method for calculating f$_{\rm H_{2}}$ with that employed by Tegmark
\etal (1997) and Yoshida \etal (2004).  As the electrons left over
from recombination are depleted, the production of H$_{2}$ slows down.
This leads to the ``asymptotic'' H$_{2}$ fraction, ${\rm f_{H_{2},
asym} ~{\tilde{\propto}}~ T^{1.52}}$, where $T$ is assumed to be equal
to the virial temperature of the halo (see Eqn. 17 of Tegmark \etal
1997).

\begin{figure}
\begin{center}
\epsfig{file=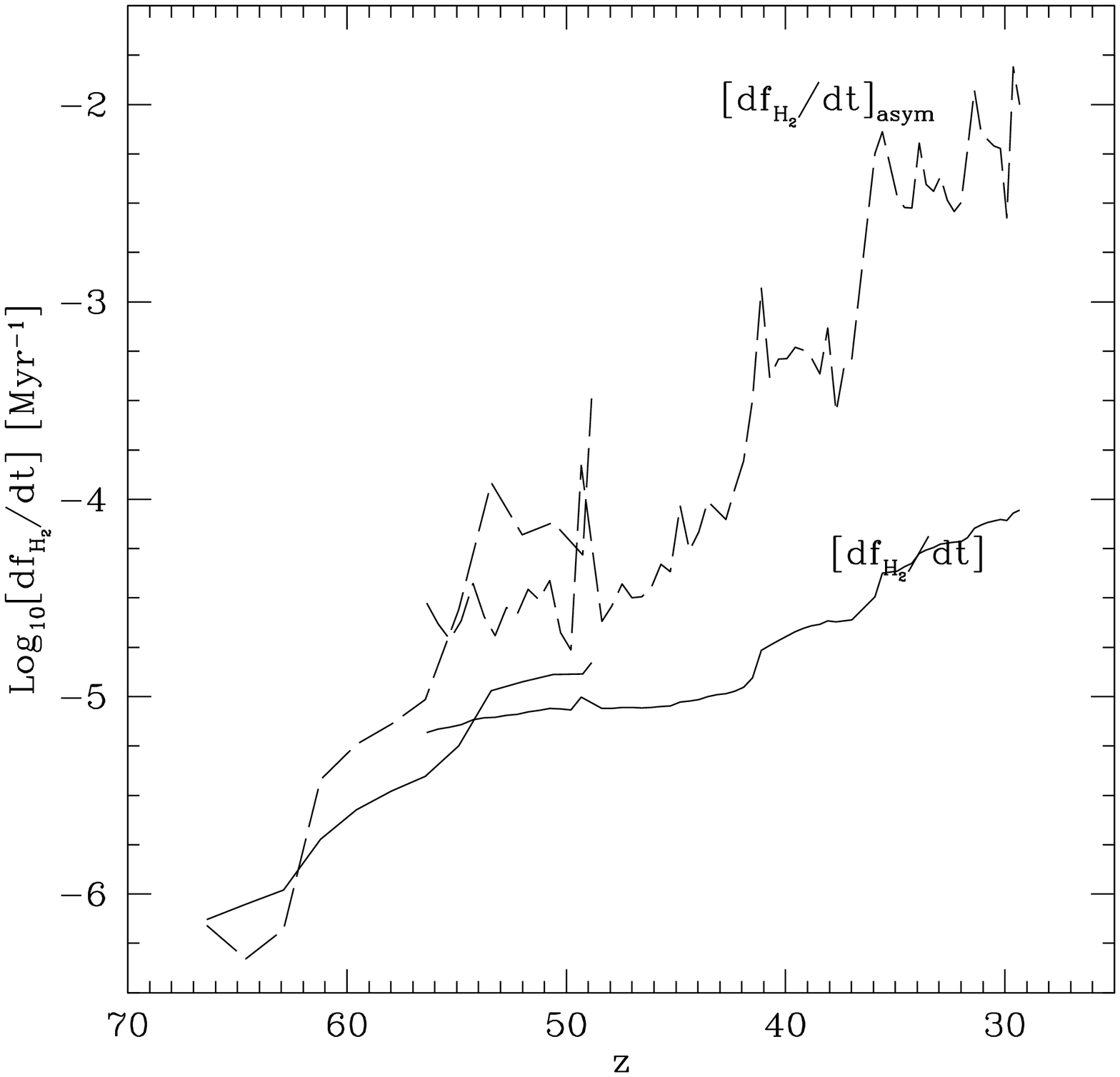, width=\hsize}
\caption{Comparison of the estimated production rate of molecular
hydrogen ${\rm df_{H_{2}}/dt}$ in a halo ({\it solid curve}) versus
${\rm [df_{H_{2}}/dt]_{asym}}$ ({\it dashed curve}), the rate of
production needed to maintain the ``asymptotic'' H$_2$ fraction.
${\rm f_{H_{2,asym}}}$ is the resulting H$_2$ abundance after H$_2$
production slows dramatically due to the depletion of catalysing free
electrons.  The growth of the halo is too rapid to maintain ${\rm
[df_{H_{2}}/dt]_{asym}}$ below ${\rm z \simeq 62}$, and so at ${\rm
z<62}$, the H$_2$ abundance of the halo will be below ${\rm
f_{H_{2,asym}}}$.  In this regime, the H$_2$ abundance is thus limited
primarily by the halo's rapid growth rather than by electron
depletion.  We thus determine the H$_{2}$ abundance from a time
integration of the H$_{2}$ production rate (see text).  The curves
that stop at redshift 49 follow simulation `R5' until its completion,
while the curves that continue to redshift 29 are for the same halo,
which is modelled at lower resolution in simulation `R4'.}
\label{dfh2crit}
\end{center}
\end{figure}

Since ${\rm T_{vir} \propto M^{2/3}}$, where M is the halo mass, it
follows that ${\rm df_{H_{2}, asym}/dt ~{\tilde{\propto}}~ dM/dt }$ as
the halo grows in mass, where ${\rm df_{H_{2}, asym}/dt}$ is the H$_2$
production rate needed to sustain ${\rm f_{H_{2}, asym}}$, which
increases as the halo grows.  The asymptotic H$_{\rm 2}$ abundance is
maintained only if the halo growth is sufficiently slow so that the
actual production rate
\begin{equation} {\rm
{df_{H_{2}} \over dt} > {df_{H_{2}, asym} \over dt}},
\end{equation}
where ${\rm df_{H_{2}}/dt}$ is the halo H$_2$ production rate.  If the
halo grows too rapidly, molecular hydrogen production will not be able
to keep up (even with the primordial abundance of free electrons) and
the cooling efficiency of the halo will be reduced.  In
Fig. \ref{dfh2crit}, we show that the H$_{\rm 2}$ production rate for
our largest halo drops below the rate required to maintain the
asymptotic H$_{\rm 2}$ abundance at ${\rm z=62}$.  This means that for
${\rm z< 62}$, f$_{\rm H_2}$ for the halo increases more slowly than
${\rm f_{H_{2}, asym}}$ because of the high halo growth rate. Similar
curves are readily produced for other high redshift haloes.  It is
thus necessary to estimate H$_{2}$ abundances by integrating ${\rm
df_{H_{2}}/dt}$ over time, rather than by simply adopting
``asymptotic'' H$_{2}$ abundances (we also include dilution effects
due to mergers as described in \S~3).  Some haloes have periods where
${\rm df_{H_{2}}/dt}$ (calculated from Eqn.~\ref{dfh2dteqn}) is larger
than ${\rm df_{H_{2}, asym}/dt}$, in which case we assume that
electron depletion limits H$_{2}$ production to ${\rm df_{H_{2},
asym}/dt}$.

\label{lastpage}


\begin{thebibliography}{}

\bibitem[]{earlygassim} Abel T., Anninos P., Zhang Y., Norman M.,
1997, NewA, 2, 181

\bibitem[]{1st} Abel T., Bryan G., Norman M. L., 2000, ApJ, 540, 39

\bibitem[]{1star} Abel T., Bryan G., Norman M. L., 2002, Science, 295, 93

\bibitem[]{anninosfh20} Anninos P., Norman M. L., 1996, ApJ, 460, 556

\bibitem[]{bbks} Bardeen J., Bond R., Kaiser N., Szalay, L., 1986, ApJ, 305, 15

\bibitem[]{bark} Barkana R., Loeb A., 2004, ApJ, 609, 474

\bibitem[]{bondeps} Bond R., Cole S., Efstathiou G., Kaiser N., 1991,
ApJ, 379, 440

\bibitem[]{bowereps} Bower R., 1991, MNRAS, 248, 332

\bibitem[]{bromm1} Bromm V., Coppi P. S., Larson R. R., 1999, ApJ, 527, L5

\bibitem[]{bromm2} Bromm V., Coppi P. S., Larson R. R., 2002, ApJ, 564, 23

\bibitem[]{brommrev} Bromm V., Larson R., 2004, ARA\&A, 42, 79

\bibitem[]{bromm3dstarform} Bromm V., Loeb A., 2004, NewA, 9, 353

\bibitem[]{bromm1stsn} Bromm V., Yoshida N., Hernquist L., 2003, ApJ,
596, L135

\bibitem[]{bruscoli} Bruscoli M., Ferrara A., Scannapieco E., 2002,
MNRAS, 330, L43

\bibitem[]{cen2} Cen R., 2003, ApJ, 591, 12

\bibitem[]{ciardih2} Ciardi B., Ferrara A., Abel T., 2000, ApJ, 533, 594

\bibitem[]{ciarditau} Ciardi B., Ferrara A., White S., 2003, MNRAS, 344, L7

\bibitem[]{ciardirev} Ciardi B., Ferrara A., 2005, Space Science Reviews, 116, 625

\bibitem{davis} Davis, M., Efstathiou, G., Frenk, C.S., White, S.D.M.,
1985, ApJ, 292, 381

\bibitem[]{draine} Draine B., Bertoldi F., 1996, ApJ, 468, 269

\bibitem[]{frailgrb} Frail D., \etal, 2001, ApJ, 562, 55

\bibitem[]{1996MNRAS...282..263}
Eke V.R., Cole S., Frenk C. S., 1996, MNRAS, 282, 263

\bibitem[]{sdssqso} Fan X., \etal, 2001, AJ, 122, 2833

\bibitem[]{gallirx} Galli D., Palla F., 1998, A\&A, 335, 403

\bibitem[]{gaoa} Gao L., Loeb A., Peebles P., White S., Jenkins A., 2004a
ApJ, 614, 17

\bibitem[]{gaob} Gao L., White S., Jenkins A., Stoehr F., Springel V.,
2004b, ApJ, 355, 819

\bibitem[]{gaoc} Gao L., De Lucia G., White S., Jenkins A., 
2004c, ApJ, 352, L1

\bibitem[]{gao} Gao L., White S. D. M., Jenkins A., Frenk C. S., 
Springel V., 2005, astro-ph/0503003, G05

\bibitem[]{glover} Glover S., Brand P., 2001, MNRAS, 321, 385

\bibitem[]{gou} Gou L., Mesazaros P., Abel T., Zhang B., 2004, ApJ, 604, 508

\bibitem[]{haimanposfeed} Haiman Z., Rees M., Loeb A., 1996, ApJ, 467, 522

\bibitem[]{haimanh2} Haiman Z., Rees M., Loeb A., 1997, ApJ, 476, 458

\bibitem[]{haimanh22} Haiman Z., Abel T., Rees M., 2000, ApJ, 534, 11

\bibitem[]{haimaintau} Haiman Z., Holder G., 2003, ApJ, 595, 1

\bibitem[]{pairsn} Heger A., Woosley S., 2002, ApJ, 567, 532

\bibitem[]{hminustoh2} Hirasawa T., 1969, Prog. Theor. Phys., 42, 523 

\bibitem[]{krecomb} Hutchins J.B., 1976, ApJ, 205 103

\bibitem[]{hizradiogrb} Ioka K., Meszaros P., 2005, ApJ, 619, 684

\bibitem[]{jenkinsmf} Jenkins A., Frenk C. S., White S. D. M., Colberg
J., Cole S., Evrard A., Couchman H., Yoshida N., 2001, MNRAS, 321, 372

\bibitem[]{kitss} Kitayama T., Susa H., Umemura M., Ikeuchi S., 2001, MNRAS, 
326, 1353

\bibitem[]{kithii} Kitayama T., Yoshida N., Susa H., Umemura M., 2004,
ApJ, 613, 631

\bibitem[]{kitsne} Kitayama T., Yoshida N., 2005, astro-ph/0505368

\bibitem[]{wmaptau} Kogut A., \etal, 2003, ApJS, 148, 161

\bibitem[]{lacey94} Lacey C., Cole S., 1994, MNRAS, 271, 676

\bibitem[]{machacek} Machacek M., Bryan G. L., Abel T., 2001, ApJ, 548, 509

\bibitem[]{machacekx} Machacek M., Bryan G. L., Abel T., 2003, MNRAS, 338,
273

\bibitem[]{mackeyh2} Mackey J., Bromm V., Hernquist L., 2003, ApJ, 586, 1

\bibitem[]{madminiqso} Madau P., Rees M., Voloteri M., Haardt F., Oh
S., 2004, ApJ, 604, 484

\bibitem[]{miralda-escude-hiz} Miralda-Escud\'{e} J., 2003, Science, 300, 1904

\bibitem[]{eps} Mo H., White S. D. M., 1996, MNRAS, 282, 347

\bibitem[]{nakaiii} Nakamura F., Umemura M., 2001, ApJ, 548, 19

\bibitem[]{nfw1} Navarro J. F., Frenk C. S., White S. D. M., 1996, ApJ, 462, 563 

\bibitem[]{nfw2} Navarro J. F., Frenk C. S., White S. D. M., 1997, ApJ, 490, 493

\bibitem[]{nav-vls} Navarro J. F., Hayashi E., Power C., Jenkins A., 
Frenk C., White S., Springel V., Stadel J., Quinn T., 2004, MNRAS, 349, 1039

\bibitem[]{nishifh2prim} Nishi R., Susa H., 1999, ApJ, 523, L103

\bibitem[]{omukainishi} Omukai K., Nishi R., 1999, ApJ, 518, 64

\bibitem[]{omukaihotpopiii} Omukai K., 2001, ApJ, 546, 635

\bibitem[]{op03} Omukai K., Palla F., 2003, ApJ, 589, 677

\bibitem[]{oshea2ndstar} O'Shea B., Abel T., Whalen D., Norman M., 2005, 
ApJ, 628, L5

\bibitem[]{panagia} Panagia N., 2003, CHJA\&A, 3, 115

\bibitem[]{peebles} Peebles P., Dicke R., 1968, ApJ, 567, 515

\bibitem[]{ps} Press W.H., Schechter P., 1974, ApJ, 187, 425

\bibitem[]{reed2003} Reed D., Gardner J., Quinn T., Stadel J., Fardal M.,
Lake G., Governato F., 2003, MNRAS, 346, 565

\bibitem[]{rees1985} Rees, M. J., 1985, MNRAS, 213P, 75

\bibitem[]{ricottih2} Ricotti M., Gnedin N., Shull M., 2002, ApJ, 575, 49

\bibitem[]{saslaw} Saslaw W., Zipoy D., 1967, Nature, 216, 976

\bibitem[]{iiiprops} Schaerer D., 2002, A\&A, 382, 28

\bibitem[]{metalssf} Schneider R., Ferrara A., Natarajan P., Omukai K., 2002, ApJ, 571, 30

\bibitem[]{sc} Schneider R., Ferrara A., Salvaterra R., Omukai K.,
Bromm V., 2003, Nature, 422, 869

\bibitem[]{cmbfast} Seljak U., Zaldarriaga M., 1996, ApJ, 469, 437

\bibitem[]{st} Sheth R., Tormen G., 1999, MNRAS, 308, 119

\bibitem[]{singh} Singh S., Ma C., 2002, ApJ, 569, 1

\bibitem[]{sokasian} Sokasian A., Yoshida N., Abel T., Hernquist L.,
Springel V., 2004, MNRAS, 350, 47

\bibitem[]{wmapn1} Spergel D., \etal, 2003, ApJS, 148, 175

\bibitem[]{gadget} Springel V., Yoshida N., White S. D. M., 2001, NewA, 6,
79

\bibitem[]{gadget2} Springel V., 2005, astro-ph/0505010

\bibitem[]{mill} Springel V., \etal, 2005, Nature, 435, 629

\bibitem[]{tegmark} Tegmark M., Silk J., Rees M., Blanchard A., Abel T., \&
Palla F., 1997, ApJ, 474, 1

\bibitem[]{ttigmreion} Theuns T., Schaye J., Zaroubi S., Kim T.,
Tzanavaris P., Carswell B., 2002, ApJ, 567, 103

\bibitem[]{1sthii} Whalen D., Abel T., Normal M., 2004, ApJ, 610, 14

\bibitem[]{whitespringel} White S. D. M., Springel V., 2000, ``Where are the first starts now?'', The First Stars, Proceedings, eds Weiss A., Abel T., \& Hill V., Springer, p. 327

\bibitem[]{wishartphotodetach} Wishart A., 1979, MNRAS, 187, 59

\bibitem[]{wiseabel} Wise J., Abel T., 2005, ApJ, 629, 615

\bibitem[]{yamamoto} Yamamoto K., Sugiyama N., Sato H., 1998, ApJ, 501, 442

\bibitem[]{yoshidavls} Yoshida N., Sheth R., Diaferio A., 2001, MNRAS, 328, 669

\bibitem[]{yoshidatrans} Yoshida N., Sugiyama N., Hernquist L., 2003, MNRAS, 
344, 481

\bibitem[]{yoshida} Yoshida N., Abel T., Hernquist L., Sugiyama N., 2003,
ApJ 592, 645

\bibitem[]{yoshida04} Yoshida N., Bromm V., Hernquist L., 2004, ApJ, 605, 579



\end{thebibliography}
\end{document}